\documentclass[12pt]{iopart}

\expandafter\let\csname equation*\endcsname\relax

\expandafter\let\csname endequation*\endcsname\relax

\usepackage{amsmath}
\usepackage{amssymb}
\usepackage{graphicx}
\usepackage{bm}
\usepackage{array}
\usepackage{dcolumn}
\usepackage{hepparticles}
\usepackage{heppennames}
\usepackage{hepnicenames}
\usepackage{epstopdf}
\usepackage{tikz} 
\usepackage{booktabs}
\usepackage{appendix}
\usepackage{cite}

\graphicspath{{/images/} }

\begin{document}

\title{An open-endcap blade trap for radial-2D ion crystals}

\author{Yuanheng Xie$^1$, Jiafeng Cui$^1$, Marissa D'Onofrio$^1$, A.J. Rasmusson$^1$, Stephen W. Howell$^2$, Philip Richerme$^{1,3}$}
\address{$^1$Indiana University Department of Physics, Bloomington, Indiana 47405, USA}
\address{$^2$Trusted Microelectronics Division, Naval Surface Warfare Center Crane, Crane, Indiana 47522, USA}
\address{$^3$Indiana University Quantum Science and Engineering Center, Bloomington, Indiana 47405, USA}

\date{\today}

\begin{abstract}
We present the design and experimental demonstration of an open-endcap radio frequency trap to confine ion crystals in the radial-two dimensional (2D) structural phase. The central axis of the trap is kept free of obstructions to allow for site-resolved imaging of ions in the 2D crystal plane, and the confining potentials are provided by four segmented blade electrodes. We discuss the design challenges, fabrication techniques, and voltage requirements for implementing this open-endcap trap. Finally, we validate its operation by confining up to 29 ions in a 2D triangular lattice, oriented such that both in-plane principal axes of the 2D crystal lie in the radial direction.
\end{abstract}
\maketitle

\section{Introduction}
Trapped ion systems are a leading platform for quantum computation and simulation due to their near-perfect state initialization and readout, long coherence times, and high-fidelity state manipulations \cite{Blatt2008Entangled,HAFFNER2008155,myerson2008high,Noek:13,Wang2021}. Among all ion trapping technologies, linear Paul traps confining one-dimensional ion chains have been the workhorse of quantum information processing experiments \cite{PhysRevLett.75.4714,Schmidt-Kaler2003Realization,Schmidt-Kaler2003How,PhysRevLett.97.220407,PhysRevLett.106.130506,PhysRevLett.103.120502,PhysRevLett.112.190502,monroe2021programmable}. Yet, many applications of trapped-ion quantum information are not well-matched to the capabilities of one-dimensional (1D) geometries. For example, the quantum simulation of interesting many-body systems such as geometrically frustrated lattices, topological materials, and spin-liquid states \cite{PhysRevLett.107.207209,Bermudez_2012,Nath_2015,richerme2016two,Yoshimura2015,balents2010spin} can all benefit from native 2D geometries.

Several parallel efforts to trap and manipulate 2D ion crystals are currently underway. For instance, 2D ion systems in microtrap arrays \cite{sterling2014fabrication,PhysRevLett.123.100504} and in Penning traps \cite{britton2012engineered,garttner2017measuring,PhysRevLett.122.053603} have made impressive progress over the last decade, but still face challenges of implementing fast quantum gates and individual addressing, respectively. To overcome such difficulties, several groups have proposed the trapping of 2D ion crystals using global potentials in standard or modified Paul traps \cite{Nath_2015,richerme2016two,Yoshimura2015,wang2015quantum}. The drawback of these schemes is the susceptibility of ions to radio frequency (rf) driven micromotion, which must be carefully considered during the trap-design stage to avoid potential effects such as rapid heating or loss of the ion crystal. Early ion-trapping work observed the confinement of 2D ion crystals in the ``lateral-2D" phase, for which the rf-driven micromotion exists along both in-plane and out-of-plane directions \cite{block2000crystalline}; later work minimized excess micromotion in this geometry \cite{https://doi.org/10.1002/qute.202000068} and cooled ions to near their motional ground state \cite{qiao2021double}. 

Additionally, 2D ion crystals have been trapped in the ``radial-2D" phase, for which the out-of-plane modes are co-aligned with the trap axis and remain micromotion-free \cite{donofrio2021radial,ivory2020paul}. In this geometry, 2D crystals were found to have long lifetimes, well-characterized vibrational modes, and low heating rates in the out-of-plane (transverse) direction, validating their use for quantum simulation experiments. The radial-2D crystals studied in \cite{donofrio2021radial} were confined in a linear Paul trap with ``needle" style endcap electrodes \cite{olmschenk2007manipulation}, which block optical access along the trap axis (perpendicular to the radial plane). In such traps, it is only possible to view the radial-2D crystal from the side. In order to achieve full site-resolved imaging and enable the possibility of individual addressing for this crystal geometry, it is necessary to develop a Paul trap with open line-of-sight along its central axis. 

In this work, we describe the development of an open-endcap linear rf trap that is capable of confining and resolving large numbers of ions in the radial-2D crystal phase. We begin in Section 2 with the design requirements for trapping and imaging radial-2D crystals as well as a simulation of our trap design. In Section 3, we discuss the trap fabrication and assembly while Section 4 covers the rf and dc electronics and voltage control. We demonstrate and characterize the performance of the open-endcap trap in Section 5 followed by concluding remarks in Section 6.

\section{Open-Endcap Trap Design}
\subsection{Linear Paul Traps}
Linear Paul traps are capable of confining ion crystals in one, two, or three dimensions \cite{donofrio2021radial,kaufmann2012precise}, but require significantly different parameters (such as trap sizes and applied voltages) to achieve each of these geometries. Thus, trap designs that have been optimized for holding 1D chains may prove incapable (or impractical) for confining ions in the radial-2D phase. In this section, we consider the requirements for the stable trapping of radial-2D ion crystals while ensuring sufficient optical access for site-resolved imaging.

The time-dependent potential provided by a linear quadrupole rf trap can be written as \cite{wineland1998experimental}
\begin{equation}
\label{eq:potential}
\begin{split}
  \Phi(\vec{r},t) &= \Phi_{dc}(\vec{r})+ \Phi_{rf}(\vec{r},t)\\
  &=\frac{\kappa U_{0}}{2z_{0}^2}(2z^2-\chi x^2-\gamma y^2) +\frac{V_{0}\cos (\Omega_{t}t)}{2d_{0}^2}(x^2-y^2)  
\end{split}
\end{equation}
where $U_{0}$ is the dc voltage, $V_{0}$ is the amplitude of an rf voltage with oscillation frequency $\Omega_{t}$, $d_{0}$ and $z_{0}$ are the radial and axial trap dimensions, and $\kappa$ is a geometric factor of order one determined by the trap electrodes. In Eq. \ref{eq:potential} we have also introduced the radial anisotropic factors $\chi$ and $\gamma$, which we experimentally choose to deviate slightly from one. (It is always required that $\chi + \gamma = 2$ to satisfy Laplace's equation). This small asymmetry breaks the degeneracy of the $x$ and $y$ radial axes, thereby preventing radial-2D crystals from rotating freely in the $xy$ plane.

Near the center of the trap, the potential may be approximated as a harmonic pseudopotential well
\begin{equation}
    \Phi(\vec{r})= \frac{1}{2}m(\omega_{x}^2x^2+\omega_{y}^2y^2+\omega_{z}^2z^2)
\end{equation}
where the secular resonance frequencies in the radial and axial directions can be written as,
\begin{equation}
\label{eq:wr}
    \omega_{r} \approx \omega_{x} \approx \omega_{y}~;~\omega_x= \sqrt{\frac{Q}{m}\left(\frac{q V_{0}}{4d_{0}^2}-\frac{\kappa \chi U_{0}}{z_{0}^2} \right)};~\omega_y= \sqrt{\frac{Q}{m}\left(\frac{q V_{0}}{4d_{0}^2}-\frac{\kappa \gamma U_{0}}{z_{0}^2} \right)}
\end{equation}
\begin{equation}
    \omega_{z}=\sqrt{\frac{Q}{m}\frac{2 \kappa U_{0}}{z_{0}^2}}
\end{equation} 
with ion charge $Q$, ion mass $m$, and the Mathieu ``$q$" parameter $q=2QV_{0}/md_{0}^2 \Omega_{t}^2$. Within this pseudopotential framework, we can account for rf-driven micromotion by expanding the ions' motion around their equilibrium positions \cite{wineland1998experimental,shen2014high}. To leading order, the coordinates of each ion varies in time as
\begin{equation}
    \vec{r}(t)=\vec{r}^{(0)}+\vec{r}^{(1)} \cos{(\Omega_{t}t)}+ \vec{r}^{(2)} \cos{(2\Omega_{t}t)}+...
\end{equation}
where $\vec{r}_{0}$ is the time-averaged ion position, $\vec{r}^{(1)}=(q_{x}\hat{x}+q_{y}\hat{y}+q_{z}\hat{z})r^{(0)}/2$ and $\vec{r}^{(2)}=(q_{x}^2\hat{x}+q_{y}^2\hat{y}+q_{z}^2\hat{z})r^{(0)}/32$ are the amplitudes of the first two micromotion terms.

\subsection{Design Considerations}
As the axial frequency $\omega_{z}$ of a Paul trap is increased from low to high while holding the radial frequencies $\omega_{r}$ fixed, a crystal of $N$ ions passes through a series of structural phase transitions: 1D chain, zig-zag, 3D, and finally, a radial-2D crystal with a triangular lattice structure \cite{donofrio2021radial,dubin1993theory}. Under the pseudopotential approximation, this radial-2D phase is achieved when the trap aspect ratio $\omega_z/\omega_r$ satisfies the condition \cite{richerme2016two,dubin1993theory,donofrio2021radial} 
\begin{equation}
\label{eq:alpha}
   \omega_{z}/ \omega_{r} > (2.264N)^\frac{1}{4}
\end{equation}
The primary design challenge for trapping crystals in the radial-2D phase is to choose the appropriate trap dimensions, voltages, and frequencies that ensure Eq. \ref{eq:alpha} is strongly obeyed for large numbers of ions, while keeping all parameters experimentally reasonable.

Several principles guide the selection of optimal trap parameters for radial-2D crystals. Satisfying Eq. \ref{eq:alpha} is most easily accomplished when $\omega_{z}$ is large, which requires large $U_0$ and/or small $z_0$. Yet, large $U_0$ and small $z_0$ have a deconfining effect in the radial direction: if the second term under the square root in Eq. \ref{eq:wr} grows too large, the ions will escape. To counter this effect, $V_0$ must also be moderately large while keeping $d_0$ small. Furthermore, since it is desirable to have small micromotion amplitudes, the trap drive frequency $\Omega_t$ should be made large to keep the Mathieu $q$ parameter small. Overall, these observations lead to a set of self-consistent design choices: small trap dimensions $d_{0}$ and $z_{0}$, large dc voltage $U_{0}$, moderately large rf voltage $V_{0}$, and relatively large trap frequency $\Omega_{t}$. For specificity, the experimental demonstration presented in Section 5 used the parameters $d_0 = 230~\mu$m, $z_0 = 200~\mu$m, $U_0 = 14.4$ V, \mbox{$V_0 = 150$ V}, and $\Omega_t = 2\pi \times 27.51$ MHz.

In addition to selecting the appropriate parameters as above, we also choose to implement a segmented-blade design for our linear Paul trap \cite{siverns2017ion, gulde2003experimental}. We consider three advantages of this trap geometry: 1) the open endcaps of the blades ensure that imaging is possible perpendicular to 2D ion plane; 2) the trap dimensions $d_0$ and $z_0$ can be made quite small to avoid unreasonably high voltages $V_0$ and $U_0$; 3) the blades can be designed such that they do not compromise the numerical aperture (NA) of the imaging optics. In our trap, we have angled the edges of the rf and dc blades to ensure that there are no obstructions to light collection using a NA = 0.28 imaging objective (Special Optics 54-17-29-369nm). 

\begin{figure}[t!]
    \centering
    \includegraphics[width=\columnwidth]{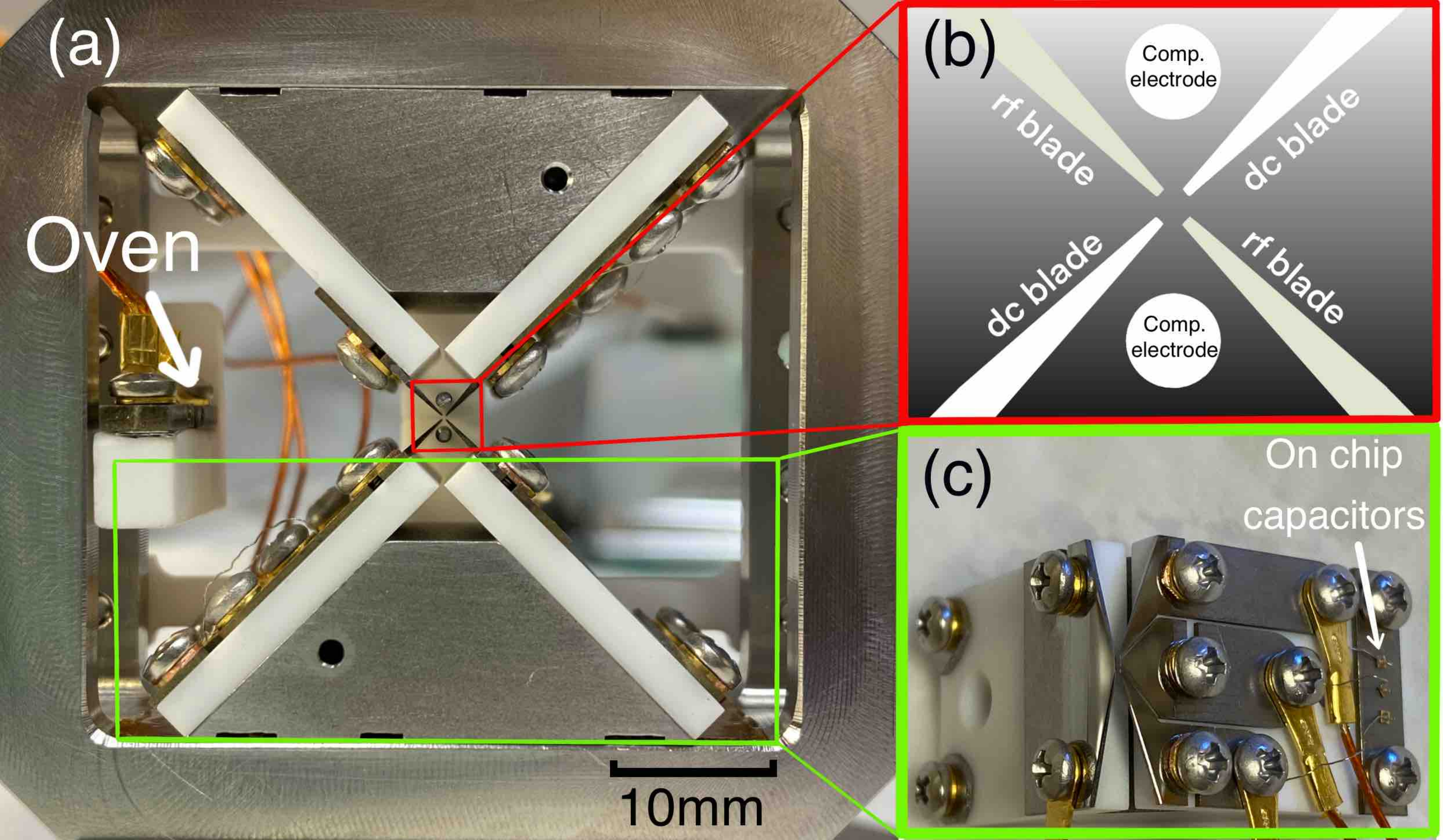}
    \caption{Assembled blade trap mounted in its vacuum chamber, taken along the imaging direction. The blades are mounted on insulating macor plates, which are fastened to a stainless steel frame and support structure (connected to ground). $^{171}$Yb and $^{174}$Yb ovens are placed to the left of the trap. (b) Sketch of the blade configuration near the trap center. Rf and segmented dc blades provide the trap potentials; two rod-style electrodes provide compensation in the vertical/horizontal directions. (c) Image of an rf blade and segmented dc blade mounted on their macor supports. Blades are machined from a 500 $\mu$m-thick piece of solid tungsten and polished after machining. On-chip capacitors (800 pF) on each dc segment provide filtering of rf pickup.}
    \label{fig:mesh1}
\end{figure}

Images of our blade-trap design are shown in Fig. \ref{fig:mesh1}. The center of the trap assembly is located 11.5 mm away the vacuum viewport to allow for a large solid angle for imaging. In this design, the dc blades are segmented into two endcaps plus one central electrode (Fig. \ref{fig:mesh1}(c)); the rf blades are continuous and of the same total length as the three dc blade segments, providing translational symmetry for the rf potential. All blades (including the rf) can be dc biased to allow for translation along all 3 principal axes. Finally, two compensation electrodes are mounted above and below the trap to provide additional voltage compensation along the vertical/horizontal directions.
\begin{figure*}[t!]
    \centering
    \includegraphics[width=\columnwidth]{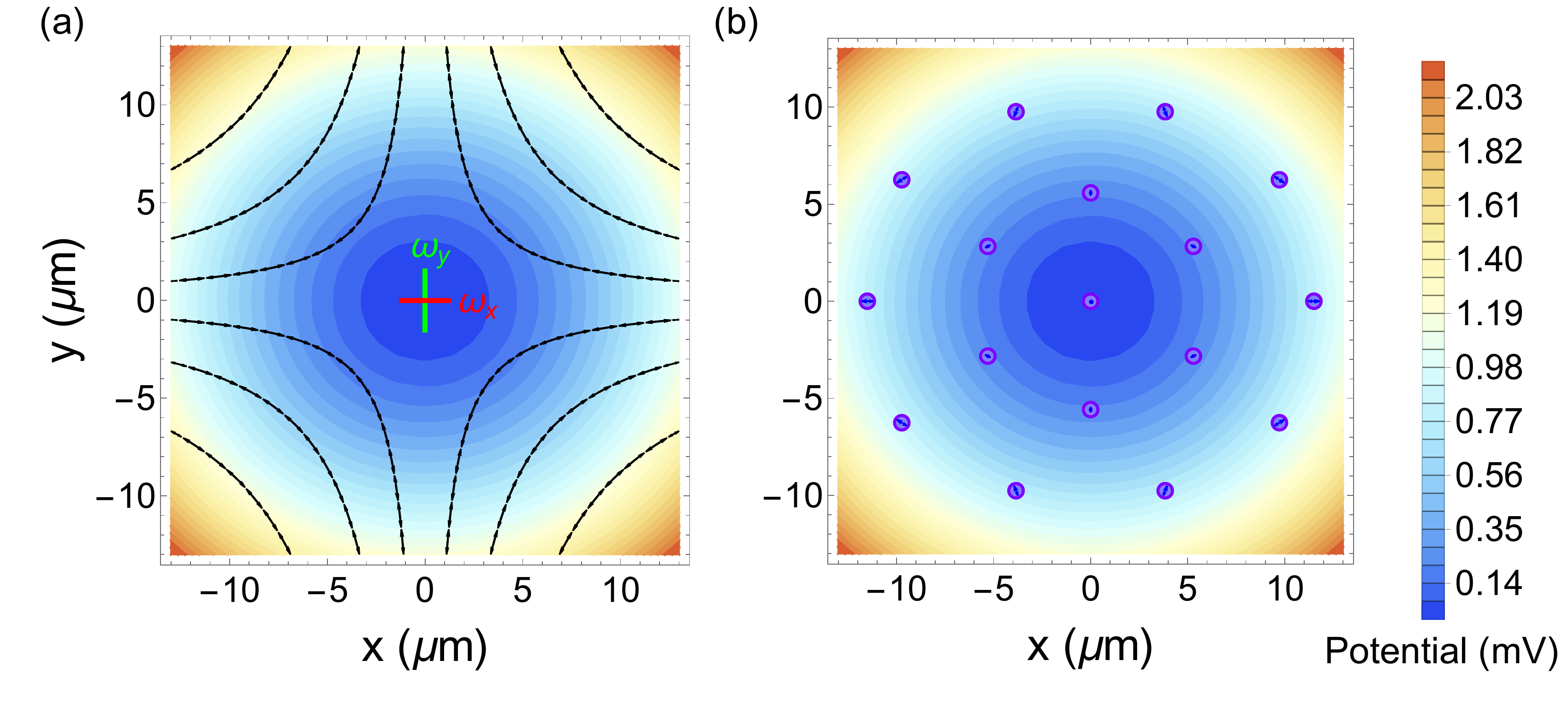}
    \caption{(a) The radial trapping potential for our open-endcap geometry, calculated using finite-element simulation methods. The two radial frequencies are made slightly non-degenerate to prevent rotation of the ion crystal. The associated electric field lines are shown in black. (b) Simulated equilibrium ion positions in for 17  $^{171}$Yb$^+$ ions in the potential of (a). Ions away from the central axes experience driven micromotion, whose amplitude can be calculated using the Floquet-Lyapunov transformation. For the 17-ion crystal, these amplitudes (shown as small arrows in (b)) are predicted to be small compared to the inter-ion spacing.}
    \label{fig:potential}
\end{figure*}
\subsection{Finite-Element Simulations}
Following the discussion above, trapping ions in the radial-2D crystal phase relies on a delicate balance between the trap geometry and the applied voltages $U_{0}$ and $V_{0}$. Equation \ref{eq:wr} shows that this balance also relies on the value of the geometric factor $\kappa$. Since the blade-style electrodes are not perfect hyperbolas (for which $\kappa=1$), it is necessary to perform numeric simulations to determine the trap secular frequencies and ensure stable trapping in the radial-2D phase.

We numerically calculate the potentials within the trap using finite-element simulations. 
First, we calculate the effective potentials $\phi_{dc}(\vec{r})$ and $\phi_{rf}(\vec{r})$ which arise from the application of 1 Volt to each individual electrode (with the others grounded). Using this set of potential basis functions, the total potential near the center of our trap can be written in the form $\Phi_{tot}(\vec{r},t)= \Phi_{rf}(\vec{r},t) +  \Phi_{dc}(\vec{r})$, where the rf contribution is given by:
\begin{eqnarray}
\nonumber
   \Phi_{rf}(\vec{r},t)&=&V_0\cos(\Omega_t t)\phi_{rf}(\vec{r}) \\
   &=&V_0\cos(\Omega_t t)(\eta _{rf}^x x^2+\eta _{rf}^y y^2+\eta _{rf}^z z^2)
   \label{eq:phirf}
\end{eqnarray}
and the dc component is:
\begin{equation}
\label{eq:phidc}
   \Phi_{dc}(\vec{r})=U_0\phi_{dc}(\vec{r})=U_0(\eta _{dc}^x x^2+ \eta _{dc}^y y^2+\eta _{dc}^z z^2)
\end{equation}
where the factors $\eta^\alpha$ in Eqs. \ref{eq:phirf} and \ref{eq:phidc} indicate the curvatures in the $\alpha$ direction for the rf and dc potentials. Comparing these equations to the form of Eq. \ref{eq:potential}, we extract the geometric factor 
$\kappa= z_{0}^2 \eta _{dc}^z$ as well as the anisotropic factors $\chi = -2z_{0}^2 \eta _{dc}^x/\kappa$ and $\gamma = -2z_{0}^2 \eta _{dc}^y/\kappa$. The resulting trap potential, along with its associated electric field, is shown in Fig. \ref{fig:potential}(a).

For a single trapped ion, the action of this simulated potential $\Phi(\vec{r},t)$ gives rise to time evolution described by the Mathieu equations
\begin{equation}
  \frac{d^2 u_{i}}{d\xi ^2} + [a_{i}-2q_{i}\cos{2\xi}]u_{i}=0
\end{equation}
where $i \in \{x,y,z\}$ and the dimensionless parameters $\xi= \Omega_{t}t/2$, $a_{i}=8QU_0\eta _{dc}^i / m\Omega_{t}^2$, $q_{i}=-4QV_0\eta_{rf}^i / m\Omega_{t}^2$. Under the pseudopotential approximation, which is valid when $a_{i} < q_{i}^2 \ll 1$, the ion secular frequencies are then defined by $\omega_{i}=\beta_i\Omega_{t}/2$, where $\beta_i\approx\sqrt{a_{i}+q_{i}^2/2} $ are the characteristic exponents of the Mathieu equation \cite{wang2015quantum,wineland1998experimental}.

Having extracted the secular trap frequencies from the finite-element simulation, we can apply the harmonic pseudopotential approximation to estimate the ion positions for large radial-2D crystals. For $N$ trapped ions, the total potential energy depends on both the trapping potential as well as the Coulomb interaction between every pair of ions:
\begin{equation}
\begin{split}
   & V(x,y,z)=\sum_{n=1}^{N} \frac{1}{2}m(\omega_{x}^2x_i^2+\omega_{y}^2y_i^2+\omega_{z}^2z_i^2) \\
    &+ \frac{Q^2}{4\pi \epsilon_{0}}\sum_{i \neq j}^N\sum_{j}^N\frac{1}{\sqrt{(x_{i}-x_{j})^2+(y_{i}-y_{j})^2+(z_{i}-z_{j})^2}}
\end{split}
\end{equation}
The equilibrium position of each ion can be found by simulating the full equations of motion with an added friction (cooling) term \cite{wang2015quantum}. The results of one such calculation, for 17 ions, are shown in Fig. \ref{fig:potential}(b). After calculating the equilibrium positions, the vibrational modes and micromotion trajectory of each ion can be calculated using the Floquet–Lyapunov transformation \cite{landa2012modes,landa2012classical} which is discussed in Appendix A. The small arrows in Fig. \ref{fig:potential}(b) show the resulting micromotion amplitude for the off-axis ions, which to first order scales linearly as the ions' radial distance from the trap center.

\section{Trap Fabrication}
\subsection{Material Selection}
Micro-fabricated, gold-coated blades are a popular choice for constructing ion trap electrodes \cite{hucul2015modular,pagano2018cryogenic}. However, the gold coating on such electrodes are often susceptible to damage from resistive heating or from large electric fields which arise during operation of the trap \cite{sterling2013increased}. For instance, tests in our lab demonstrated that the large rf voltages required for creating the radial-2D crystal phase quickly led to melting and evaporation of the gold layer. 

To ensure more robust performance in the presence of large rf voltages, we fabricated our electrodes from solid tungsten. Tungsten is an easily available, strong, and low resistivity metal that has been used in a variety of earlier rf traps \cite{PhysRevLett.97.103007,Zalivako2019Nonselective,Cruz2005Field,Arrington2013,Ray2014}. Compared with more common metals (such as stainless steel), we consider tungsten advantageous for our trap since its low resistivity will limit blade heating and any associated vacuum pressure increases when large rf voltages are applied.

\subsection{Blade Fabrication and Assembly}
\begin{figure}[h]
    \centering
    \includegraphics[width=\columnwidth]{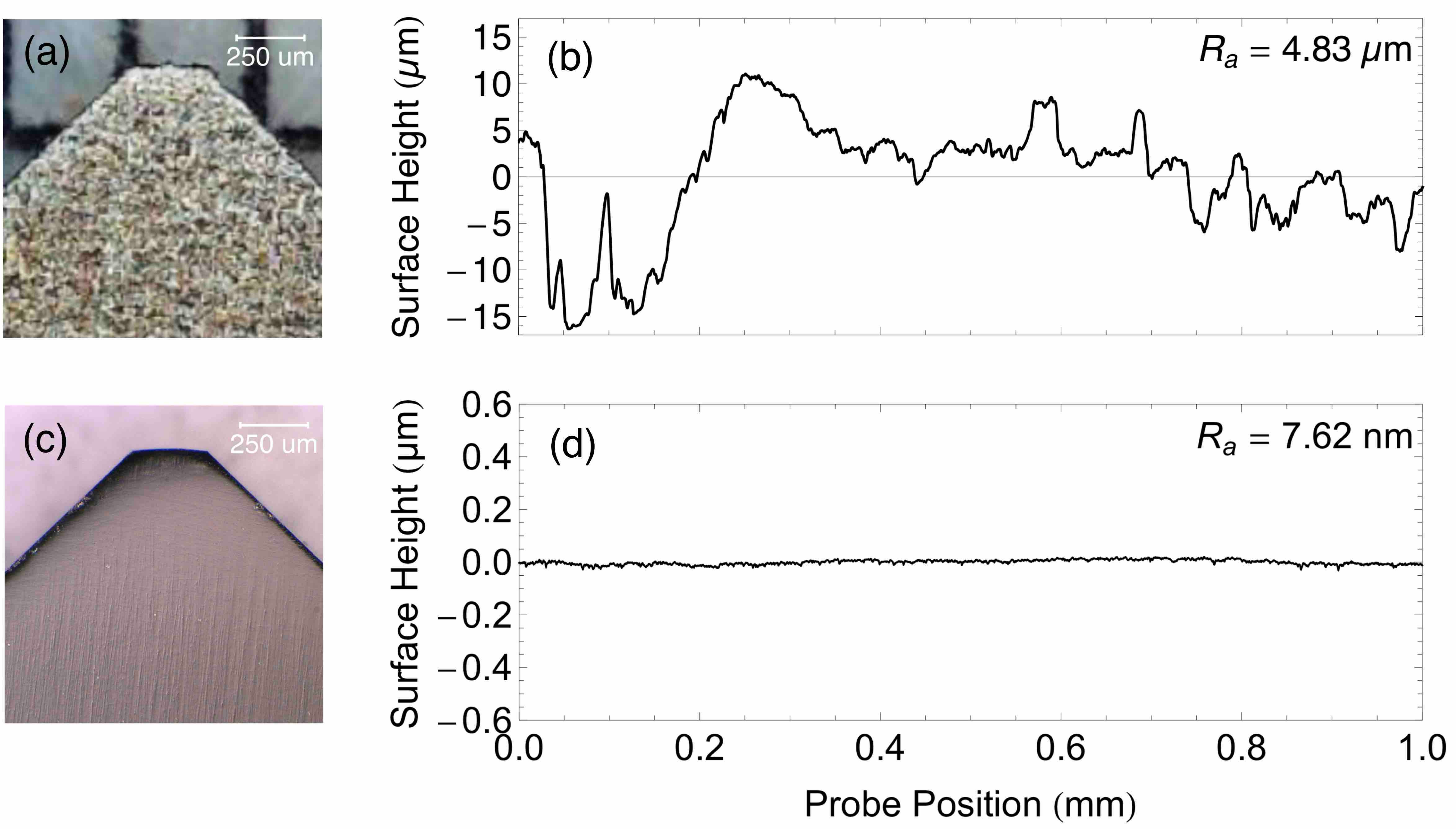}
    \caption{(a) Image of a blade electrode directly after wire-EDM machining. (b) Using a stylus profilometer near the tip of the blade, we characterize the average surface roughness $R_a$. (c) After electropolishing and hand polishing, the blade has a smooth mirror-like surface. (d) The surface roughness of the polished blade is reduced by nearly three orders of magnitude compared to the unprocessed blade.}
    \label{fig:blades}
\end{figure}
The blade electrodes are fabricated from a sheet of 500 $\mu$m-thick pure tungsten using a wire-EDM (Electrical Discharge Machining) process. This technique allows for fairly complex electrode geometries (such as the segmented dc blades) to be machined to within $\sim10~\mu$m tolerances. In our design, the three dc blade segments are each 300 $\mu$m long and separated by a 50 $\mu$m gap; the rf blades have a total length of 1 mm. The final processed tip thickness is 100 $\mu$m for all blades; however, for reasons explained below, we target an initial 300 $\mu$m tip thickness during wire-EDM machining.

For sintered materials like tungsten, the exposed surface following wire-EDM processing can be markedly rough. This can be problematic for trapped-ion systems, since there is evidence that large surface roughness could significantly affect motional heating rates \cite{PhysRevA.94.013418,hite2013surface, PhysRevLett.109.103001, brownnutt2015ion}. In addition, rough electrode surfaces could increase unwanted laser beam reflections, increasing the background light collected by the imaging optics. Fig. \ref{fig:blades}(a) shows an image of a blade electrode just after wire-EDM machining. Using a stylus profilometer (Bruker DektakXT), we characterized the arithmetic average surface roughness $R_a$ of this blade to be approximately $5~\mu$m over a 1 mm region near the tip (Fig. \ref{fig:blades}(b)).

To smooth the tungsten blade surface after machining, we implement a two-stage polishing process. First, we use self-terminated electrochemical etching to remove the largest surface features \cite{Wang2016}. The blade is immersed into sodium hydroxide solution (NaOH, 400 ml of 2 mol/L) and connected to the cathode of a power supply (10 V, 1.5 A) for 2 minutes of etching. This process lowers the surface roughness to $R_a < 1~\mu$m and reduces the tip thickness from 300 $\mu$m to $\sim100~\mu$m. Following this electropolishing stage, the electrodes are hand-polished using $3~\mu$m, $1~\mu$m, and $0.3~\mu$m stages of Aluminum Oxide polishing paper. Fig. \ref{fig:blades}(c) shows an image of the blade electrode after processing. As measured by a profilometer, the surface roughness is reduced from $R_a \approx 5~\mu$m to \mbox{$R_a \approx 8$ nm} after these polishing processes (Fig. \ref{fig:blades}(d)).

Following machining and polishing of the blades, the trap is assembled in a clean room to avoid dust contamination. As shown in Fig. \ref{fig:mesh1}(a), the blades are mounted on macor plates fastened to trapezoidal stainless-steel blocks. The dc segments are hand-aligned under a microscope to keep a $50~\mu$m gap between segments, and the rf blade is mounted parallel to the dc segments with a gap of $280~\mu$m (Fig. \ref{fig:mesh1}(c)). Two assembled triangular blocks are placed in a stainless-steel frame in a vertex-to-vertex orientation, with a vertical gap between blades of $300~\mu$m. All dc electrodes are mechanically connected to gold-plated lugs, which are crimped to kapton-coated wire and connected to a Sub-C 9-pin feedthrough. The rf blades and atomic ovens connect with separate high-power electrical feedthroughs. To reach UHV pressures, the vacuum system was initially pumped to $10^{-7}$ Torr, then baked for two weeks at $200^{\circ}$C; the final pressure of the chamber at room temperature is below $10^{-11} $ Torr. 

\section{Electronics and Voltage Control}
\subsection{Helical Resonator}
\begin{figure}[h]
    \centering
    \includegraphics[width=\columnwidth]{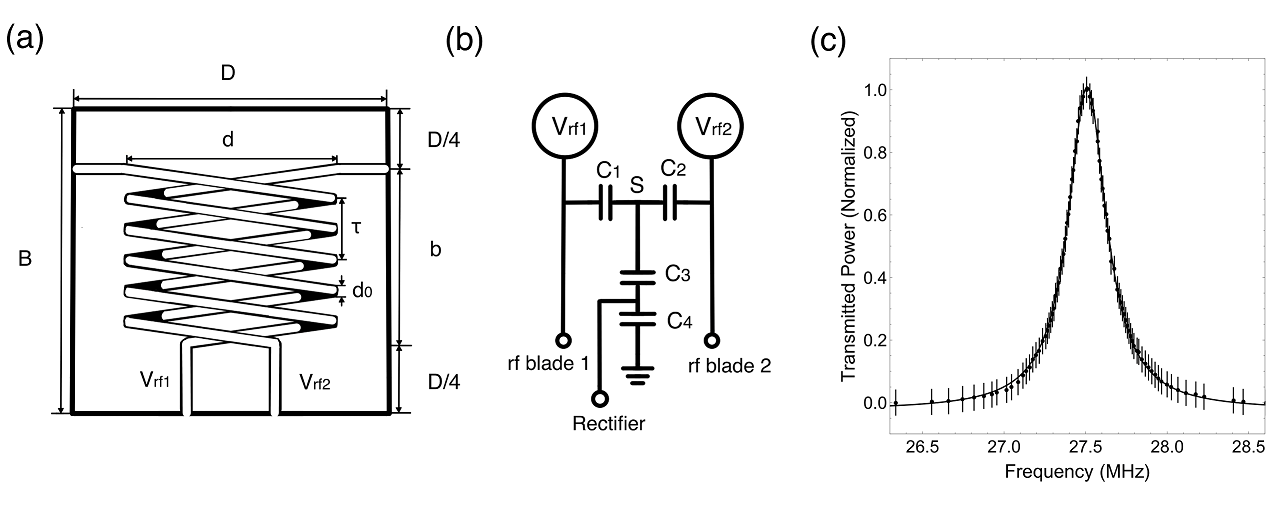}
    \caption{(a) Sketch of the two-coil resonator design. The shield diameter $D$ = 102 mm, shield height $B$ = 101 mm, coil diameter $d$ = 64 mm, coil height $b$ = 68 mm, winding pitch $\tau$ = 10 mm, and the coil wire diameter $d_0$ = 2.5 mm provides a $\Omega_{t} = 2\pi \times 27.51$ MHz drive frequency when connected to a trap with capacitance $C_{trap}=10$ pF. (b) Schematics of the voltage sampler and capacitive divider. $C_1$ and $C_2$ are placed across the outputs to balance $V_{\mathrm{rf}1}$ and $V_{\mathrm{rf}2}$; $C_3$ and $C_4$ form the capacitive divider. $C_1$, $C_2$, $C_3$ and $C_4$ are on the same circuit board housed within the resonator cylinder. The voltage-reduced sampling signal is sent out through a BNC connector. The rectifier circuit is right next to the BNC connector to avoid electromagnetic interference. (c) We measure a resonator $Q$ factor of $\approx 100$ by sampling the output of the capacitive divider as a function of rf input frequency.}
    \label{fig:Helical}
\end{figure}

Ions confined by a Paul trap require a stable, high voltage, and low noise rf potential. A helical resonator allows impedance matching between the rf source and the ion trap, amplifying voltage while filtering noise injected into the system \cite{Siverns2012}. We opt to build a two-coil resonator, since this allows for independent dc biasing of the rf blades so that the trap may be compensated in all directions. In order to construct a resonator to operate at a desired frequency, we first measure the capacitance of the connection wire and ion trap $C_{\mathrm{trap}}$ at the trap feedthrough. Once these are known, we build the shield and helical resonator coils following the procedure outlined in \cite{Siverns2012}. A cross-sectional drawing of the two-coil resonator is shown in Fig. \ref{fig:Helical}(a) along with our chosen design parameters. Under these conditions, the resulting resonant frequency is $\Omega_{t} = 2\pi \times 27.51$ MHz when connected to our blade trap. 

Due to our implementation of a two-coil resonator, we designed the circuit shown in Fig. \ref{fig:Helical}(b) to appropriately sample the transmitted rf voltage. To begin, we use two capacitors $C_1$ and $C_2$ (KEMET, SMD Comm X5R series, 10 $\mu$F) to bridge between the resonator's two outputs $V_{\mathrm{rf}1}$ and $V_{\mathrm{rf}2}$. With this configuration, we can accomplish two goals. First, combining the two rf outputs with capacitors balances any potential phase differences caused by mechanical asymmetry of the resonator. Second, the potential at the point $S$ is the average value of the outputs which can be used as a voltage probing point for sampling.

A capacitive divider connected to point $S$ is used to scale down the high rf-voltage for sampling. The divider consists of two high voltage-tolerance capacitors, $C_3$ (AVX Corporation, SQ series, 0.2 $\mu$F) and $C_4$ (AVX Corporation, UQ series, 20 $\mu$F). This combination picks off 1$\%$ of the high voltage signal (down to the $\sim 1$ V range) so that the later rectifier design requirements are less stringent. We measured the rf pickup signal as a function of input rf frequency, from which we determined the $Q$ factor of the resonator to be $\sim 100$ (as shown in Fig. \ref{fig:Helical}(c)).

\subsection{rf Locking and Stability}

The fidelity of quantum operations within ion traps is sensitive to fluctuations in the rf frequency, which may be driven by noise in the input rf amplifier, mechanical vibrations of the resonator, and temperature drifts (to name a few sources). Since the trap secular frequencies depend on the ratio of $V_0/\Omega_t$ (Eq. \ref{eq:wr}), and since the rf drive frequency $\Omega_t$ is typically well-stabilized at the rf source, active rf amplitude stabilization is a crucial tool for keeping the trap secular frequency consistent. In our case, we actively stabilize the rf voltage amplitude following the techniques outlined in \cite{Johnson2016Active}. 

\begin{figure}[t]
    \centering
    \includegraphics[width=\columnwidth]{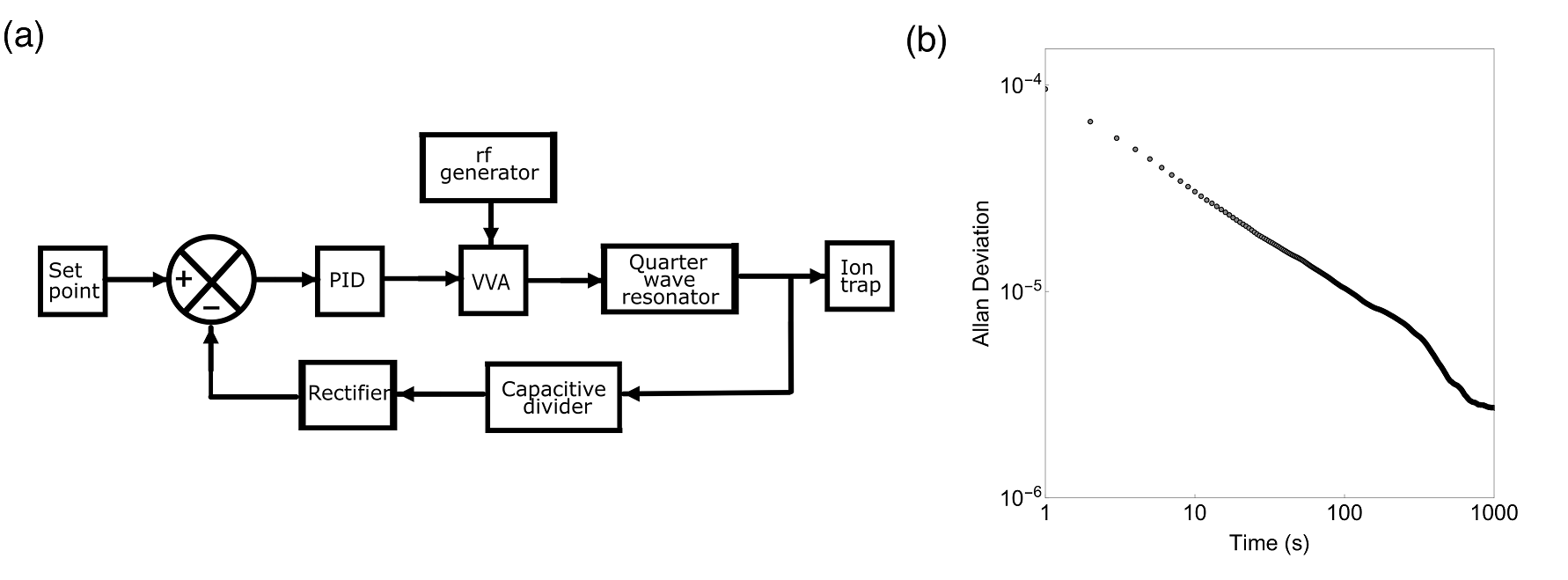}
    \caption{ (a) Servo loop block diagram for active stabilization of the rf voltage amplitude. VVA: Voltage Variable Attenuator, PID: proportional-integral-derivative controller. (b) Allan deviation of the rf signal amplitude during operation of the servo loop in (a).}
    \label{fig:rfdrive}
\end{figure}

Our servo loop block diagram is shown in Fig. \ref{fig:rfdrive}(a). The rf generator produces a signal at frequency $\Omega_t = 2\pi \times 27.51$ MHz and power -8 dBm, which passes through a voltage variable attenuator (VVA) and is amplified before being sent to the helical resonator. The picked-off signal from the resonator passes through a rectifier circuit and is fed as the input of a closed proportional–integral–derivative (PID) loop. The rf amplitude is thus stabilized with respect to the set point value. 

We performed long-term monitoring of the dc signal after the rectifier when the servo is engaged. These measurements represent the scale of rf amplitude fluctuations over time. We find that the Allan deviation of rf amplitude scales with time $\tau$ as $\approx 1/\sqrt{\tau}$; at 1000 s, the relative stability is $2.74 \pm 0.04\times 10^{-6}$, which translates to a \mbox{$\sim 30$ Hz} rms fluctuation of the radial secular frequencies.

\subsection{dc Control}

In addition to rf control, we desire stable, low noise voltages applied to the dc trap electrodes. We apply these voltages using static dc power supplies (Matsusada R4G series) that can output 0-120 V with 1 mVrms ripple. To prevent noise pickup from various ancillary electronic devices, each dc channel is externally RC-filtered before being connected to a 9-pin feedthrough at the trap vacuum chamber.

\begin{figure}[t!]
    \centering
    \includegraphics[width=\columnwidth]{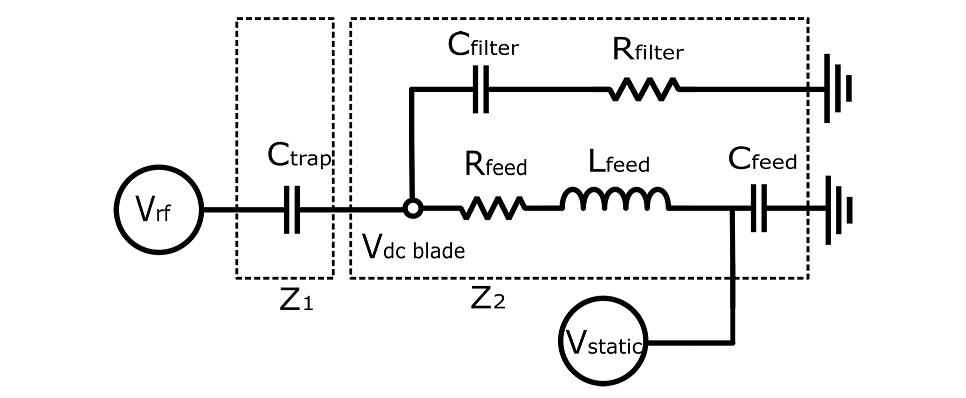}
    \caption{Dc circuit diagram of our ion trap. The in-vacuum filter is designed to reduce the rf pickup on the static dc blades. See text for details and component values.
    }
    \label{fig:dccircuit}
\end{figure}

Given the small dimensions of our blade trap, one additional concern is unwanted rf pickup on the dc blades. To mitigate this effect, each dc blade segment is wirebonded to an 800 pF capacitor to shunt rf pickup to ground (see image in Fig. \ref{fig:mesh1}(c)). To model the effective in-vacuum circuit, we treat the ion trap as a capacitor ($C_{trap}$) and consider the contributions from the on-trap filter elements and vacuum feedthroughs, as shown in Fig. \ref{fig:dccircuit}. Using this model, we can estimate the rf pickup on the static dc blades by first calculating the complex impedance
\begin{equation}
     Z_2=
     \left(\frac{1}{Z_{C,filter}+Z_{R,filter}}+\frac{1}{Z_{C,feed}+Z_{R,feed}+Z_{L,feed}}\right)^{-1}
\end{equation}
where for our system, $Z_{R,feed}\approx R_{feed}\ll 1~ \Omega$,  $|Z_{C,feed}|=\frac{1}{\Omega_{t}C_{feed}}\approx 1.8~\Omega$ and  $|Z_{L,feed}|=\Omega_t L_{feed}\approx 52~\Omega$. The resistance of the filter $R_{filter} \ll 1~\Omega$, which is negligible compared with the capacitive filter impedance $|Z_{C,filter}|=\frac{1}{\Omega_{t}C_{filter}}=7.2~\Omega$. Thus, we estimate the impedance $|Z_{2}|=6.4~\Omega$. The measured trap capacitance of 10 pF leads to an impedance $|Z_{1}|=600~\Omega$ at our trap drive frequency. Therefore, the estimated rf pickup on the dc blades is then $\frac{|Z_{2}|}{|Z_{1}|+|Z_{2}|} V_{RF}=0.01V_{RF}$. We note that in the absence of the on-trap filter capacitors, the rf pickup on the dc blades would be approximately a factor of 8 larger.

\section{Experimental Demonstration}
\subsection{Laser access and internal states}
\begin{figure}[t]
    \centering
    \includegraphics[width=12cm]{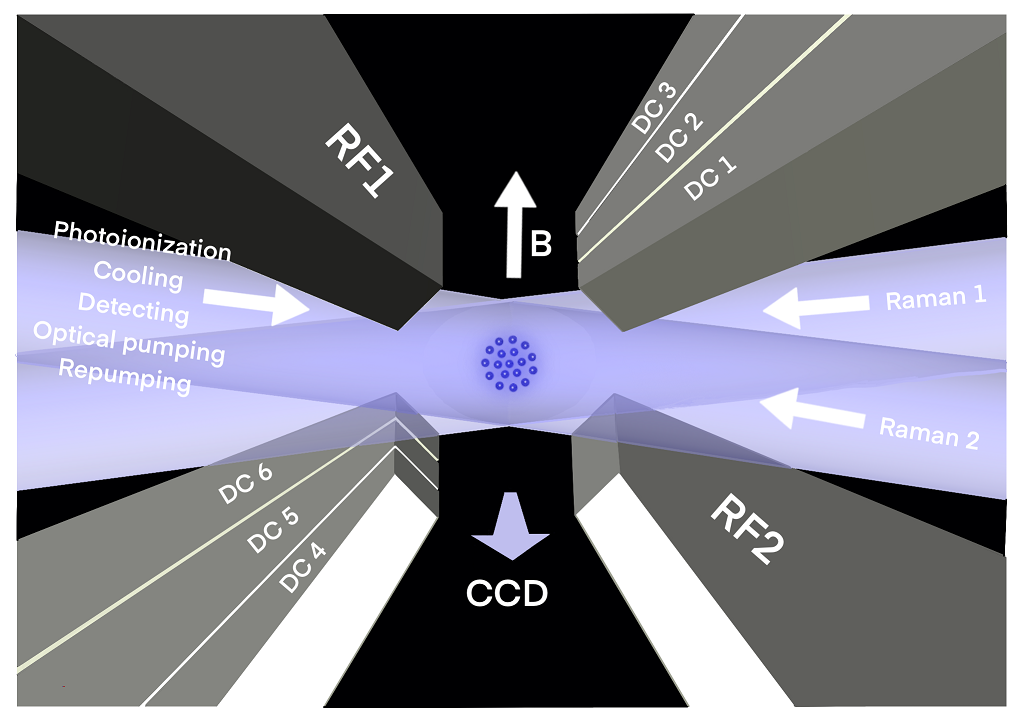}
    \caption{Concept drawing of the trap and laser beam configurations for photoionization (399 nm), cooling, optical pumping, and detection (369.5 nm), repumping (935nm), and two-photon Raman transitions (355nm). The CCD camera faces the crystal plane and the magnetic field is oriented vertically. Oscillating voltages on electrodes RF1 and RF2 provide the radial confinement, while static voltages applied to electrodes DC1, DC3, DC4, and DC6 provide axial confinement.}
    \label{fig:trapsketch}
\end{figure}

Our blade trap and vacuum system have been designed to ensure sufficient optical access for state preparation, manipulation, and measurement. As shown in Fig. \ref{fig:trapsketch}, ${}^{171}$Yb$^+$ ions are loaded into the trap via photo-ionization of neutral $^{171}$Yb using 399 nm and 369.5 nm light. Ions are Doppler cooled by 369.5 nm light that is $\approx 10$ MHz red-detuned of the ${}^{2}S_{1/2}- {}^{2}P_{1/2}$ transitions, and co-aligned with the 399 nm beam. Additional wavelength components near 369.5 nm are used for optical pumping and detection of the qubit state, while co-aligned light at 935 nm is used to repump population out of the metastable $^2D_{3/2}$ state \cite{PhysRevA.76.052314}. A 5 G magnetic field along the vertical direction breaks the degeneracy of the hyperfine triplet. Finally, two-photon stimulated Raman transitions for quantum state manipulation are driven by shining two 355 nm beams such that their wave vector difference $\Delta \vec{k}$ is aligned along the transverse direction of the radial-2D crystal (which is the axial direction of the trap). 

\subsection{Confinement and imaging of radial-2D crystals}

\begin{figure}[t]
    \centering
    \includegraphics[width=\columnwidth]{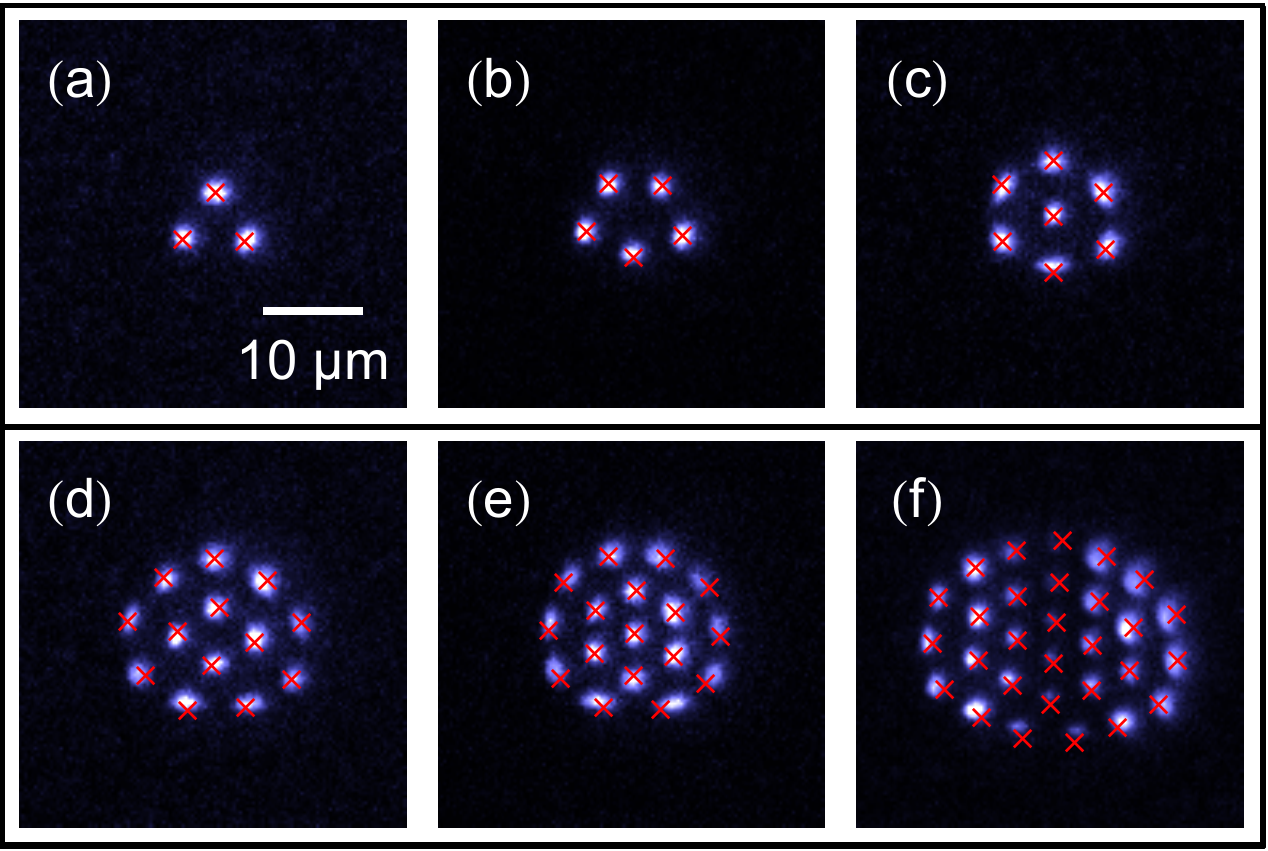}
    \caption{CCD images of crystals with 3, 5, 7, 13, 17, and 29 ions trapped in the radial-2D crystal phase, with measured center-of-mass frequencies $\omega_x= 2\pi\times 0.416$ MHz, $\omega_y= 2\pi\times 0.446$ MHz, $\omega_z=  2\pi\times 1.124$ MHz. Red crosses show the ion positions predicted under the pseudopotential approximation}
    \label{fig:2dions}
\end{figure}

Ions may be confined in radial-2D crystals once the trap secular frequencies satisfy the inequality in Eq. \ref{eq:alpha}. To create this ion geometry experimentally, we load the desired number of ions at low axial frequency $\omega_z$, then increase the endcap voltages (DC1, DC3, DC4, DC6) to push the ions into the radial-2D phase. In practice, imperfect electrode fabrication, trap misalignments, and stray electric fields could cause ion heating during the transitions through different structural phases. To avoid losing ions, and to minimize any excess micromotion, we compensate by applying small bias voltages to blade segments as needed. Once ions are in the 2D regime and Doppler-cooled to milliKelvin temperatures, they form a Wigner crystal as the system minimizes its configuration energy. As shown in Fig. \ref{fig:2dions}, the final crystal geometry takes the form of a triangular lattice in the radial plane.

The ion positions can be predicted under the pseudopotential approximation once the trap frequencies are experimentally known. To measure the ion secular frequencies for a set of applied trap voltages, we inject an additional small rf voltage on the trap electrode DC 3 following Doppler cooling. This electrode is chosen since its contribution to the electric field at the ion has components along the $\hat{x},\hat{y}$, and $\hat{z}$ directions. If the injected rf drive is in resonance with the ion's oscillation frequency, the ion will absorb energy and heat up, decreasing its fluorescence when probed with a detection laser beam \cite{nagerl1998coherent}. Once the frequencies are determined, we use the procedure outlined in Sec 2.3 to predict the ion positions in the radial-2D crystal; results are shown as red crosses in Fig. \ref{fig:2dions}.

More complex processes are involved when laser cooling ion crystals in two- and three-dimensions as compared to the one-dimensional case. Ions away from the trap center experience micromotion, which leads to Doppler-shifted cooling transitions which depend on the micromotion amplitude at each ion position. For large crystals, this micromotion-induced Doppler shift can lead to dramatically different cooling rates for a crystal's outermost ions as compared to the inner ions \cite{devoe1989role}. Optimum Doppler cooling is often found further red-detuned than the typical single-ion detuning, which may result in relatively decreased fluorescence for the innermost ions (as seen in Fig. \ref{fig:2dions}(f)). For very large crystals, it may ultimately prove necessary to introduce multi-tone Doppler cooling to frequency-address ions at different radii, or to power-broaden the resonant transition as suggested in \cite{devoe1989role}.
 
\subsection{Ion Trajectory Analysis}

In radial-2D crystals, ions located far from the origin experience the largest amplitude of micromotion. Following the process outlined in Sec 2.3 (and Appendix A), we calculate that the maximum radial micromotion amplitude in a 17-ion radial-2D crystal is $< 650$ nm, which is small compared to the $\sim 1~\mu$m diffraction-limited spot size of our imaging system as well as the 5 $\mu$m inter-ion spacing. The micromotion along the axial direction is calculated to be negligible due to the small Mathieu $q_{z}$ parameter; this was confirmed in prior measurements using the ``needle trap" geometry, where the radial-2D crystal was imaged from the side \cite{donofrio2021radial}. Since the out-of-plane axial modes remain micromotion-free in this geometry, these modes will be preferable for performing future quantum simulation experiments.

Close inspection of the ion trajectories in Fig. \ref{fig:2dions} reveals a convex curvature, rather than the concave curvature which would be expected from driven micromotion (see calculation in Fig. \ref{fig:potential}(b)). We attribute this result to thermally-driven, small-angle rotations of the ion crystal. Consider, for instance, a radial-2D crystal in a perfectly-symmetric potential with degenerate radial secular frequencies. The crystal will be free to rotate with no energy penalty; when imaged on a CCD camera, the ions will appear as concentric rings. Although in our trap this degeneracy is explicitly broken, residual thermal energy in the crystal may still induce small azimuthal oscillations. We calculate that the ion excursions observed in Fig. \ref{fig:2dions}(e)-(f) are consistent with crystal temperatures of only $\approx 20$ mK. We anticipate that this effect can be reduced by further breaking the degeneracy between the radial secular frequencies, by introducing multi-tone Doppler cooling \cite{devoe1989role}, or by applying sub-Doppler cooling techniques such as resolved sideband cooling \cite{wineland1998experimental,diedrichlaser1989} or Electromagnetically-Induced Transparency (EIT) cooling \cite{qiao2021double,lechner2016electromagnetically,feng2020efficient}.

\section{Conclusion}
Radial-2D crystals hold great promise as a platform for quantum simulation of exotic many-body materials but require carefully-designed rf traps to realize a robust implementation. In this work, we have presented the design of an open-endcap blade trap which allows for both stable confinement of radial-2D crystals as well as site-resolved imaging of the triangular 2D lattice. The electrode geometry has been chosen so that the required trap potentials are accessible using reasonable laboratory voltages, and the use of tungsten as an electrode material limits potential damage from resistive heating or voltage flashover. Furthermore, we have implemented rf and dc electronics which lead to stable and low-noise operation of the trap.

Our experimental observation of up to 29 ions in radial-2D arrays, imaged perpendicularly to the crystal plane, paves the way for 2D quantum simulation experiments. Previous work has established that such radial-2D crystals are long-lived, and that the out-of-plane vibrational modes remain cold and isolated from rf-heating effects \cite{donofrio2021radial}. The open-endcap trap technology presented here, which allows for imaging along the trap axis, represents the final step needed for quantum simulation experiments to proceed. In future work, we anticipate implementing individual-ion addressing in this open-endcap design, which will further expand the classes of quantum materials that may be simulated using radial-2D crystals in linear rf traps.

\ack
This work was supported by the U.S. Department of Energy, Office of Science, Basic Energy Sciences, under Award $\#$DE-SC0020343. The IU Quantum Science and Engineering Center is supported by the Office of the IU Bloomington Vice Provost for Research through its Emerging Areas of Research program. We are grateful to Thomas Dean Smith in the IU Mechanical Instrument Services Shop for his assistance with fabricating the trap electrodes. We also thank Dr. John Holaday for assistance with optical imaging, and Drs. Tyler Smith and Jonathan Dilger for assistance with roughness measurements and helpful discussions.

\section*{References}
\bibliographystyle{prsty}
\bibliography{main}{}

\appendix
\section{Dynamic Solution of Ion Motion}
In this section, we precisely solve the normal modes and micromotion of N ions in a 2D crystal by the Floquet–Lyapunov transformation \cite{landa2012modes,landa2012classical}. The potential energy of the ions in our Paul trap can be written as
\begin{equation}
\begin{split}
    V & =V_{trap}+V_{coulomb} \\ & =\displaystyle\sum_{i}^{n}\frac{1}{2}(\Lambda_xx_i^2+\Lambda_y y_i^2+\Lambda_z z_i^2)+ \displaystyle\sum_{i\neq j}\frac{1}{2}\frac{q^2}{4\pi \epsilon_0}\| r_i-r_j \|^{-1} 
\end{split}
\end{equation}
where $r_i=\{x_i,y_i,z_i\}$ is the vector coordinate of ion $i$, and the time-dependent trapping terms are given by 
\begin{equation}
   \Lambda_\alpha=B_\alpha+A_\alpha\cos (\Omega_t t),\ \alpha\in\{x,y,z\}
\end{equation}
$A_\alpha,B_\alpha$ represent the real trap electric potential coefficients. In Sec 2.3 we calculated the secular frequencies under the pseudopotential approximation, which may be expressed as
\begin{equation}
    V_{pseudo}=\frac{1}{2}m\displaystyle\sum_{i}^{n}(\omega_x^2x_i^2+\omega_y^2y_i^2+\omega_z^2z_i^2)
\end{equation}

The total potential energy could then be written as
\begin{equation}
\begin{split}
\label{eq:potential1}
    V&=V_1+V_2\\
    &=(V_{pseudo}+V_{coulomb})+(V_{trap}-V_{pseudo})
\end{split}
\end{equation}
Treating $V_{2}$ as the perturbation, we expand the time-dependent positions $\{R_{i,\alpha}(t)\}$ around the minimum-configuration locations $\{R_{i,\alpha}^0\}=(x_{1}^{(0)},y_{1}^{(0)},z_{1}^{(0)},$\dots$ ,x_{N}^{(0)},y_{N}^{(0)},z_{N}^{(0)})$ that are obtained from the secular part of $V_1=V_{pseudo}+V_{coulomb}$. The time-dependent positions can then be written in terms of the normal modes $S_j$ by setting
\begin{equation}
\label{eq:postions}
   R_{i,\alpha}(t)=R_{i,\alpha}^0 + r_{i,\alpha}=R_{i,\alpha}^0 + \displaystyle\sum_{j}^{3N}\Gamma_{i,j}S_j(t)
\end{equation}
where  $\Gamma_{i,j}$ are the matrix elements of the normal mode vectors, with rows indexed by the $N$ ions $i$ in the three directions $\alpha$, and columns indexed by the $3N$ normal modes $j$. 

We then plug Eq. \ref{eq:postions} into Eq. \ref{eq:potential1}, write the potential in terms of the normal modes, and keep the first two terms:
\begin{equation}
\begin{split}  
\label{eq:potentials2}
  V=\frac{1}{2}\vec{S}^T\Lambda \vec{S}+\displaystyle\sum_{i,\alpha}^N(\Lambda_\alpha-\frac{1}{2}m\omega_\alpha^2)(R_{i,\alpha}^0 + \displaystyle\sum_{j}^{3N}\Gamma_{i,j}S_j)^2+\dots\\
  \approx \frac{1}{2}\vec{S}^T\Lambda \vec{S}+((\vec{R}^0)^T+\vec{S}^T\Gamma)(W_1+W_2cos\Omega t)(\vec{R}^0+\Gamma^T\vec{S})
\end{split}
\end{equation} 
where $\Lambda =$ diag$\{\Omega_{i\alpha}^2\}$, $W_1=$ diag$\{B_\alpha-\frac{1}{2}m\omega_\alpha^2\}$, $W_2=$ diag$\{A_\alpha\}$, and $\Omega_i$ is the $i$th normal frequency in $\alpha$ direction. The linearized equation of motion derived from Eq. \ref{eq:potentials2} is 
\begin{equation}
\label{eq:EOM}
  mS''+(\Lambda +J)\cdot S+P+(L+Y\cdot S)\cos\Omega t=0 
\end{equation}
where 
\begin{equation}
\begin{split}
  \vec{P}=\Gamma \cdot W_1\cdot \vec{R}^0+(\vec{R}^0)^T\cdot W_1 \Gamma^T\\\
  \vec{L}=\Gamma \cdot W_2\cdot \vec{R}^0+(\vec{R}^0)^T\cdot W_2 \Gamma^T\\
  J=\Gamma \cdot W_1\cdot \Gamma^T+(\Gamma \cdot W_1\cdot \Gamma^T)^T\\
  Y=\Gamma \cdot W_2\cdot \Gamma^T+(\Gamma \cdot W_2\cdot \Gamma^T)^T
\end{split}
\end{equation}
Let
\begin{equation}
\begin{aligned}
  A&=(\Lambda+J)\frac{4}{\Omega^2 m}\\
  Q&=-\frac{1}{2}Y\frac{4}{\Omega^2 m}\\
  \vec{G}&=-\vec{P}\frac{4}{\Omega^2 m}\\
  \vec{F}&=-\frac{1}{2}\vec{L}\frac{4}{\Omega^2 m}\\
\end{aligned}
\end{equation}

We then have a simplified inhomogeneous Mathieu Matrix Equation from Eq.\ref{eq:EOM}
\begin{equation}
\label{eq:eom}
  \vec{S}''+(A-2Q\cos\Omega t)\cdot \vec{S}=\vec{G}+2\vec{F}\cos\Omega t
\end{equation}
where $F$ and $G$ are $3N$-component constant vectors. We assign the basic $\pi$ periodic solution $\vec{S}=\displaystyle\sum_{-\infty}^{\infty}\vec{B}_{2n} e^{i(2n)t}$
in the equations of motion (Eq. \ref{eq:eom}) to obtain
\begin{equation}
\begin{split}
 (A-4n^2 &) \vec{B}_{2n}-Q(\vec{B}_{2n-2}+\vec{B}_{2n+2}) = \\
  &\vec{G}\delta _{1,n}+\vec{F}(\delta _{n,1}+\delta _{n,-1})  
\end{split}
\end{equation}
By defining $C_{2n}=A-4n^2$ and using $B_{2n}=B_{-2n}$, we can write infinite recursion relations for $ \vec{B}_{2n}$,
\begin{equation}
\label{eq:G}
  A\vec{B}_0-2Q\vec{B}_2=\vec{G}
\end{equation}
\begin{equation}
\label{eq:F}
  C_2\vec{B}_2-Q(\vec{B}_0+\vec{B}_4) =\vec{F}
\end{equation}
\begin{equation}
\label{eq:C2n}
  C_{2n}\vec{B}_{2n}-Q(\vec{B}_{2n-2}+\vec{B}_{2n+2})=0,(n \geq 2)
\end{equation}
Eq. \ref{eq:C2n} immediately gives a recursion relation in the form of Eq. \ref{eq:eom}, which allows us to get the infinite inversions expression
\begin{equation}
\label{eq:B4}
   \vec{B}_4=T_2Q\vec{B}_2 
\end{equation}
where
\begin{equation}
    T_2=[C_4-Q[C_6-Q[C_8-...]^{-1}Q]^{-1}Q]^{-1}
\end{equation}
Substituting Eq. \ref{eq:B4} into Eq. \ref{eq:G} and \ref{eq:F} we obtain the linear system
\begin{equation}
\left( \begin{array}{cc} A & -2Q \\
-Q & R_2-QT_2Q \end{array} \right)
\left( \begin{array}{c} \vec{B}_0\\ \vec{B}_2 \end{array} \right)=\left( \begin{array}{c} \vec{G}\\ \vec{F} \end{array} \right)
\end{equation}
which can be solved to find the coefficients of the normal modes $\vec{S}$, and the micromotion terms $\vec{r}=\Gamma^T \cdot \vec{S}$.

\end{document}